\documentclass{PoS}

\PoS{PoS(LAT2005)359}

\title{Finite-volume effects in moving frames} 

\ShortTitle{Finite-volume Effects on two hadron system in moving frames}

\author{\speaker{Changhoan Kim}\\
        University of Southampton, School of Physics and Astronomy,\\
	Highfield, Southampton, SO17 1BJ, United Kingdom\\
        E-mail: \email{chateau@phys.soton.ac.uk}}

\author{Chris T. Sachrajda \\
        University of Southampton, School of Physics and Astronomy,\\
	Highfield, Southampton, SO17 1BJ, United Kingdom \\ 
        E-mail: \email{cts@phys.soton.ac.uk}}

\author{Stephen R. Sharpe \\
        University of Washington, Department of Physics,\\
	Seattle, WA-98195-1560, USA \\ 
        E-mail: \email{sharpe@phys.washington.edu}}

\abstract{
We determine the quantization condition for the energy levels of
two interacting particles in a finite box in a ``moving frame'', 
i.e. one in which the total momentum of pions is non-zero.
This condition is valid up to corrections which fall exponentially
withe the box size, and holds only below the
inelastic threshold. It is derived using field theoretic methods,
using a generalization of previous summation formulae
relating sums and integrals over momenta.
The result agrees with that obtained earlier by
Rummakainen and Gottlieb using a relativistic quantum mechanical
approach.
Technically, we expand 
the finite-volume four-point Green function in terms of the 
infinite-volume Bethe-Salpeter kernel,
and determine the position of the poles.
The final result is written 
in terms of the two-pion scattering phase shift. 
Our result can be used to facilitate the determination of the
scattering phase shift, and can be used to generalize the
Lellouch-L\"uscher formula relating finite-volume two-particle
matrix elements to those in infinite volume.
}

\FullConference{XXIIIrd International Symposium on Lattice Field Theory\\
		 25-30 July 2005\\
		 Trinity College, Dublin, Ireland}
\usepackage{amssymb,amsmath}
\usepackage{epsfig}
\usepackage{psfrag}
\newcommand{\be}{\begin{equation}}
\newcommand{\ee}{\end{equation}}
\newcommand{\bea}{\begin{equation}\begin{array}}
\newcommand{\eea}{\end{array}\end{equation}}
\newcommand{\bdm}{\begin{displaymath}}
\newcommand{\edm}{\end{displaymath}}
\newcommand{\rba}{\begin{array}}
\newcommand{\rea}{\end{array}}
\newcommand{\bi}{\begin{itemize}}
\newcommand{\ei}{\end{itemize}}
\newcommand{\bver}{\begin{verbatim}}
\newcommand{\everbatim}{\end{verbatim}}

\begin{document}
\section{Introduction}
Lattice simulations are necessarily performed in a finite volume.
Unlike the exponentially small errors in lattice evaluation of
 hadronic masses or matrix elements with at most a single
hadron in the external states, the errors  when two
(or more) hadrons are present in external states the finite-volume
 effects decrease more slowly, as powers of the
box size, $L$, and need to be understood in order to obtain physical
quantities with good precision.
The theory of such effects has been fully developed for two particles
in their rest frame, i.e. with total momentum $\vec P=0$.
The spectrum of such states was worked out in
refs.~\cite{ml2,ml3,ml4} and the finite volume corrections to
the matrix elements were obtained in refs.~\cite{ll,lmst}.

In this note, we report on a determination of the spectrum of 
 two-particle states with total momentum $\vec P\ne 0$,
which we call a {\em moving frame}.\footnote{A generalisation of the results
for the spectrum of such states to a moving frame was proposed
some time ago in ref.~\cite{rg}. We address this in sec.~\ref{sec:con}}
Full details are given in ref.~\cite{kss}.\footnote{%
The results have been also obtained in ref.~\cite{cky} using a
different method.}

There are several reasons why this extension is important, e.g.
the use of a moving frame obviates
the need for vacuum subtractions in the isoscalar channel, which is
 otherwise under poor statistical control. Further applications
are discussed in refs.~\cite{rg,kss}.

The energy levels of two-particle system are determined 
from the position of poles of 
corresponding correlation function.
By expanding the correlation function in terms of Bethe-Salpeter kernel,
 we can locate the terms which potentially generate power corrections.
With the aid of summation formulae, 
we can calculate these corrections up to exponentially small errors.
After rearranging and resumming we deduce the {\em quantization condition}, 
whose solutions generate the location of poles.
This condition depends on the scattering phase shift and the box size.
Thus, we can 
 determine the scattering phase shift from a lattice calculation of spectrum. 
We sketch the derivation of the summation formulae in sec.~\ref{sec:sumfor} and
of the quantization condition in sec.~\ref{sec:qcmf}.

\section{Summation Formulae}\label{sec:sumfor}
  Finite-volume effects for two-hadron states originate from the
  difference between the sums over the discrete momenta in a finite
  volume and the corresponding integrals over the continuous spectrum
  in infinite volume.  To determine the relation between
  them, we start from the Poisson summation formula,
\begin{equation}\label{eq:poisson_f}
\frac{1}{L^3}\sum_{\vec{k}}\,g(\vec{k}\,)=
\sum_{\vec l}\int\frac{d^3k}{(2\pi)^3} e^{i\, L \,\vec
l\cdot \vec k} g(\vec{k}\,)\,
\approx
\int\frac{d^3k}{(2\pi)^3}\,g(\vec{k}\,) + O(e^{-L})
\end{equation}
where the summation on the left-hand-side is over 
all integer values of $\vec{n}=(n_1,n_2,n_3)$, with $\vec{k}=(2\pi/L)\,\vec{n}$.
The approximation is valid when
 the Fourier transform of $g(\vec k,)$, $\tilde{g}(\vec{r}\,)$, is
 non-singular, and is either contained in a finite
spatial region or decreases exponentially as $|\vec{r}\,|\to\infty$.
We note that functions $g(\vec k\,)$ with these properties have
no singularities for real $\vec k$,
and fall off fast enough at $|\vec k|\to\infty$ that the integrals
in eq.~(\ref{eq:poisson_f}) converge.

However, the finite-volume corrections for two-hadron correlators with non-zero total momentum are contained in summations of the form 
\begin{equation}\label{eq:movingsum}
S(q^\ast)\equiv\frac{1}{L^3}\sum_{\vec{k}}\
\frac{\omega_k^\ast}{\omega_k}\ \frac{f(\vec{k}^\ast)} {q^{\ast\,
2}-k^{\ast\, 2}}\,.
\end{equation}
where we assume that $q^{*2}$ is such
that there is no term in the sum with $k^{*2}\equiv |\vec{k^*}\,|^2=q^{*2}$ and that
$f(\vec{k^*}\,)$ has the properties discussed above.
Note that the summation is over
the moving frame momenta $\vec{k}=(2\pi/L)\vec{n}$,
with $\vec{n}$ being a vector of integers but the
summand is written in terms of the centre-of-mass momenta $\vec{k}^\ast$
using the Lorentz transformation of eq.(\ref{eq:lorentz}). It
is also convenient, as will be apparent shortly, to pull the Jacobian
$\omega_k^\ast/\omega_k$ out of the function $f(\vec{k}^\ast\,)$.

The manifest singularity at $k^2=q^2$ forbids the immediate application
 of the eq.~(\ref{eq:poisson_f}).
In order to avoid this difficulty, 
 we expand $f$ in terms of spherical harmonics\footnote{%
Here, we used $\sqrt{4\pi}\, Y_{lm}$, which simplifies the
normalization for $l=0$.} 
and subtract from $S(q^*)$ 
a function which is specially chosen  to cancel the singularity: 
\begin{eqnarray}
\lefteqn{\frac{1}{L^3} \sum_{\vec{k}}\
J\ \frac{f_{lm}(k^\ast)-f_{lm}
(q^\ast)e^{\alpha(q^{\ast\,2}-k^{\ast\,2})}}
{q^{\ast\,2}-k^{\ast\,2}}\
k^{\ast\,l}\sqrt{4\pi}\,Y_{lm}(\theta^\ast,\phi^\ast)}
\nonumber\\
&&=\int\frac{d^{\,3}k}{(2\pi)^3}\,J 
\frac{f_{lm}(k^\ast)-f_{lm}(q^\ast)
e^{\alpha(q^{\ast\,2}-k^{\ast\,2})}} {q^{\ast\,2}-k^{\ast\,2}}\
k^{\ast\,l}\sqrt{4\pi}\,Y_{lm}(\theta^\ast,\phi^\ast)
\label{eq:intmv}\\
&&=\int\frac{d^{\,3}k^\ast}{(2\pi)^3}\
\frac{f_{lm}(k^\ast)-f_{lm}(q^\ast)e^{\alpha(q^{\ast\,2}-k^{\ast\,2})}}
{q^{\ast\,2}-k^{\ast\,2}}\
k^{\ast\,l}\sqrt{4\pi}\,Y_{lm}(\theta^\ast,\phi^\ast)\,.\label{eq:intcom}
\end{eqnarray}
where the Jacobian factor $\omega_k^\ast/\omega_k$ corresponds to the
change of integration variables from the laboratory-frame momenta
$\vec k$ to the centre-of-mass frame momenta $\vec{k}^\ast$. 
This relation is valid up to terms which are exponentially small in the volume.
The exponential
factors $\exp[\alpha(q^{*2}-k^{*2})]$ (with $\alpha>0$) are
included so that the subtraction does not introduce ultraviolet
divergences.
By rearranging terms, the required summation formulae for given spherical harmonics is
\footnote{This equality holds up to exponentially small corrections. 
From now on, this will not be stated explicitly.}
\begin{equation}\label{eq:sflm}
S_{lm}(q^*)=\delta_{l,0}\ {\cal
P}\int\frac{d^{\,3}k^*}{(2\pi)^3}\,\frac{f_{00}(k^*)}{q^{*2}-k^{*2}}
+f_{lm}(q^*) c^P_{lm}(q^{*\,2})\,,
\end{equation}
where
\begin{equation}\label{eq:zlmdef}
c^P_{lm}(q^{*\,2})=\frac{1}{L^3}\sum_{\vec{k}}
\frac{e^{\alpha(q^{*2}-k^{*2})}}
{q^{*2}-k^{*2}}\,k^{*l}\,\sqrt{4\pi}\,Y_{lm}(\theta^*,\phi^*)
-
\delta_{l,0}\ {\cal
P}\int\frac{d^{\,3}k^*}{(2\pi)^3}\frac{e^{\alpha(q^{*2}-k^{*2})}}
{q^{*2}-k^{*2}}\,.
\end{equation}
Although the integrand in eq.\,(\ref{eq:intcom}) has no pole
at $k=q$, in eq.\,(\ref{eq:sflm}) we separate it into two
terms each of which does have such a pole. For consistency, the
two terms need to be regulated in the same way and the principal
value prescription, denoted ${\cal P}$, is a natural choice.
Finally, we gather all the spherical harmonics and then the summation formula is
\begin{equation}
S(q^\ast)={\cal P}\int\frac{d^{\,3}k^\ast}{(2\pi)^3}\
\frac{f(\vec{k}^\ast)} {q^{\ast\,2}-k^{\ast\,2}}+
\sum_{l=0}^\infty\,\sum_{m=-l}^l\,
f_{lm}(q^\ast)\, c^P_{lm}(q^{\ast\,2})\,.
\end{equation}
\section{Quantization Condition in Moving Frames}\label{sec:qcmf}
The two-pion spectrum in finite volume can be extracted from the
 correlation functions of composite operators:
\begin{equation}\label{eq:CPdef}
C_{\vec{P}}(t)=
\langle\,0\,|\,\sigma_{\vec{P}}(t)\,\sigma^\dagger(\vec{0},0)\,|\,0\rangle,
\quad\quad
\widetilde{C}_{\vec{P}}(E)=\int dt\, e^{-iEt}\,C_{\vec{P}}(t)\,.
\end{equation}
(with time ordering implicit),
where $\sigma(\vec{x},t)$ is an interpolating operator for
two-pion states and $\sigma_{\vec{P}}(t)$ is its spatial Fourier
transform. 
We will determine the quantization condition
 by identifying the position of poles of $\widetilde{C}_{\vec{P}}(E)$.  

\begin{figure}[t]
\begin{center}
\epsfxsize=\hsize
\epsfbox{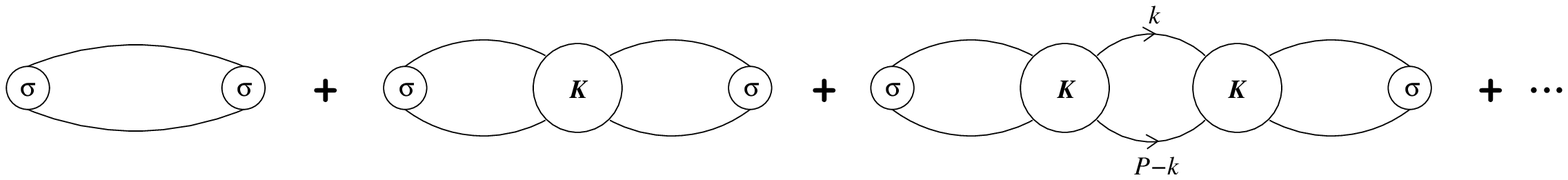}
\end{center}
\caption{Diagrammatic expansion of the correlator $\widetilde C_{\vec P}(E)$.
Propagators are fully dressed and normalized to unity
on shell. $K$ is the amputated two-particle irreducible
four-particle correlation function. The circles at the ends
represent the operator $\sigma$, renormalized by two factors
of $\sqrt{Z}$.}
\label{fig:CP}
\end{figure}

The correlation function $\widetilde C_{\vec P}$ can be expressed in
terms of the Bethe-Salpeter kernel $K$ through the series shown in fig.\,\ref{fig:CP}.
Since we choose $E$ to lie below the
four-pion threshold, there are no intermediate states with four or
more pions and the finite-volume effects in $K$
 are exponentially suppressed~\cite{ml2,lmst}.
The same is true of the dressed single particle propagators~\cite{ml1}.
The only power-law volume corrections arise through the
the two pion loops, and we now turn to an analysis of these corrections.
The generic loop integration/summation appearing in fig.~\ref{fig:CP}
is of the form
\footnote{
Although we phrase our discussion in Minkowski space, we note that
this same object may be obtained from the Euclidean space correlators
calculated in lattice simulations
by analytic continuation to imaginary Euclidean energy
(which is the approach used in ref.\,\cite{ml2}).}
\begin{equation}
I\equiv\frac{1}{L^3}\sum_{\vec{k}}\,\int
\frac{dk_0}{2\pi}\frac{f(k_0,\vec{k})}{(k^2-m^2+i\varepsilon)
((P-k)^2-m^2+i\varepsilon)}\,
\end{equation}
where $k=(k_0,\vec{k})$ and $P=(E,\vec{P})$ are four vectors and
we have left out a factor of $i^2=-1$ from the propagators which
will be accounted for later.
The only properties of
$f$ that we need are, first, that it has no singularities for real
$\vec k$ (which holds given our kinematical constraint on $E$),
and, second, that its ultraviolet behaviour is such as to render
the integral and sum convergent.

To simplify the pole structure we first perform the $k_0$ integration.
We choose to close the contour of integration so as to
pick up the ``particle'' contribution from the first pole and
the ``anti-particle'' contribution from the second:
\begin{equation}
I=-i\frac{1}{L^3}\sum_{\vec{k}}\left\{\frac{f(\omega_k,\vec{k})}{2\omega_k
((E-\omega_k)^2-\omega_{Pk}^2)}+
\frac{f(E+\omega_{Pk},\vec{k})}{2\omega_{Pk}
((E+\omega_{Pk})^2-\omega_k^2)}
\right\}\,,\label{eq:iresult}
\end{equation}
 where $\omega_k=\sqrt{\vec{k}^2+m^2}$ and $\omega_{Pk}=\sqrt{(\vec{P}-\vec{k})^2+m^2}$.

For the kinematic region of interest, $0<E^2-P^2<16 m^2$, it is
straightforward to show that the only singularity in $I$ is
the explicit pole in the first term inside the braces in
eq.\,(\ref{eq:iresult}), which occurs at those values of $E$ for which there is
a term in the summation with $\omega_k+\omega_{Pk}=E$.
Since this singularity leads to finite-volume corrections which
decrease like powers of the volume, we need to examine the first term in more detail.

With Lorentz transformed variables,
\begin{equation}\label{eq:lorentz}
k^\ast_\parallel=\gamma(k_\parallel-\beta\omega_k),\quad
\tilde{k}^\ast_\perp=\tilde{k}_\perp\quad\textrm{and}\quad
\omega_k^\ast=\sqrt{k^{\ast\,2}+m^2}=\gamma(\omega_k-\beta
k_\parallel)\,.
\end{equation}
we rewrite the first term (which we call $I_1$) in the form
\begin{equation}
I_1=-i\frac{1}{L^3}\frac{1}{E^\ast}\sum_{\vec{k}}\frac{1}{2\omega_k}
\frac{f^\ast(\vec{k}^\ast)}
{E^\ast-2\omega_k^\ast}=\,
-i\frac{1}{L^3}\frac{1}{2E^\ast}\sum_{\vec{k}}\frac{\omega_k^\ast}{\omega_k}
\,\frac{f^\ast(\vec{k}^\ast)}{q^{\ast\,2}-k^{\ast\,2}}\,\frac
{E^\ast+2\omega_k^\ast}{4\omega_k^\ast}\,,
\end{equation}
where $f^\ast$ is the function $f$ rewritten in terms of the
centre-of-mass variables.
We now apply the summation formulae from sec.~\ref{sec:sumfor},
\begin{equation}
I_1=-i\frac{1}{2E^\ast}\,{\cal P}\int\frac{d^3k^\ast}{(2\pi)^3}
\,\frac{f^\ast(\vec{k}^\ast)}{q^{\ast\,2}-k^{\ast\,2}}\,\frac
{E^\ast+2\omega_k^\ast}{4\omega_k^\ast}\,-\,\frac{i}{2E^\ast}
\sum_{l=0}^{\infty}\sum_{m=-l}^{l}
 f^\ast_{lm}(q^\ast)\,c^{P}_{lm}(q^{\ast\,2})\,.
\label{eq:i1v1}\end{equation}

\begin{figure}[t]
\begin{center}
\epsfxsize=\hsize
\epsfbox{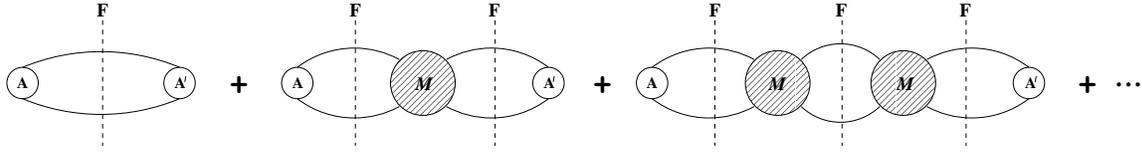}
\end{center}
\caption{Contributions to the volume dependent part
of the $\sigma$ correlator, $\widetilde C_{\vec P}^{FV}$.
The notation is as in fig.~\protect\ref{fig:CP} except that the
filled circles represent the full scattering amplitude, $M$,
given by a geometric sum of any number of
insertions of the kernel $K$, and the vertical dashed lines
indicate that the on-shell finite volume part, $I_{FV}$, has
been used for the loop integral. The quantities $A$ and $A'$ are
defined in the text.}
\label{fig:CPFV}
\end{figure}

In order to isolate the finite-volume correction, 
 we replace the principal-value integral in
eq.\,(\ref{eq:i1v1}) by the corresponding integral with the
Feynman $i\varepsilon$ prescription in the propagator and
a ``delta-function'' term:
\begin{equation}
I_1=-i\frac{1}{2E^\ast}\,\int\frac{d^3k^\ast}{(2\pi)^3}
\,\frac{f^\ast(\vec{k}^\ast)}{q^{\ast\,2}-k^{\ast\,2}+i\varepsilon}\,\frac
{E^\ast+2\omega_k^\ast}{4\omega_k^\ast}
+\frac{q^\ast\,f^\ast_{00}(q^\ast)}{8\pi E^\ast} -\,
\frac{i}{2E^\ast}
\sum_{l=0}^{\infty}\sum_{m=-l}^{l}
 f^\ast_{lm}(q^\ast)\,c^{P}_{lm}(q^{\ast\,2})
\,. \nonumber
\label{eq:i1v2}\end{equation}
Note that the ``delta-function'' term picks out the $l=0$ part of $f^\ast$.
Observing that the first term in eq.\,(\ref{eq:i1v2}) is
exactly the infinite volume expression for $I_1$ in Minkowski space
(after retracing the steps in the derivation above),
we arrive at the finite-volume correction of loop integrals, 
\begin{equation*}
I=I_\infty+I_{FV}\,, \quad
I_{FV} = \left\{\frac{q^\ast\,f^\ast_{00}(q^\ast)}{8\pi
E^\ast} -\,\frac{i}{2E^\ast}\,\sum_{l=0}^{\infty}\sum_{m=-l}^{l}
f^\ast_{lm}(q^\ast)\, c^{P}_{lm}(q^{\ast\,2})\right\}\,.
\label{eq:IFV}
\end{equation*}
Since each loop integral (sum) in fig.~\ref{fig:CP} contains an
infinite volume part and the finite volume correction, we can
rearrange the expansion according to the number of insertions
of the latter. Ignoring the zero insertion term, which is
irrelevant to the pole structure of the correlation function, we can
represent the series as in fig.~\ref{fig:CPFV}, leading to the
general result:
\begin{equation}
\widetilde{C}_{\vec P}^{FV}(E) = -A'\; F\; A + A'\;F\;(iM/2)\;F \; A + \dots
= - A'\; F\; \frac{1}{1 + i M F/2}\; A
\,.\label{eq:CPFVres}
\end{equation}
Here we have taken into account the factor of
$i^2$ dropped from the loop in the previous section, as well as
symmetry factors
of $1/2$ arising from the identical nature of the particles.
$A$ and $A'$ represent the overlap of $\sigma$ with asymptotic two-pion states 
with definite angular momentum.
$F$ is a kinematic factor obtained from $I_{FV}$:
\begin{equation}
F_{l_1,m_1;\,l_2,m_2} = \frac{q^\ast}{8\pi E^\ast}
\left( \delta_{l_1 l_2} \delta_{m_1 m_2}
-i\frac{4\pi}{q^\ast}\,
\sum_{l=0}^{\infty} \sum_{m=-l}^{l}
\frac{\sqrt{4\pi}}{q^{\ast\,l}} c^{P}_{lm}(q^{\ast\,2})
\int d\Omega^\ast \, Y^\ast_{l_1,m_1}
\,Y^\ast_{l,m}\, Y_{l_2,m_2} \right) \,,
\end{equation}
and $M$ is the on-shell scattering amplitude:
\begin{equation}
M_{l_1,m_1;\,l_2,m_2} = \delta_{l_1 l_2}\; \delta_{m_1 m_2}
\frac{16\pi E^\ast}{q^\ast}\,
\frac{\left(\exp[{2i\delta_{l_1}(q^\ast)}]-1\right)}{2 i}
\,,\label{eq:Mlm}
\end{equation}
Now, we can determine the final quantization condition: 
\begin{equation}
\mathrm{det}(1 + i M F/2) = 0
\,.\label{eq:det}
\end{equation}
Solving this after truncation of the partial wave expansion leads
to results which can be shown to be equivalent to those of ref.~\cite{rg}.

\section{Conclusion}\label{sec:con}
We have provided a field theoretic derivation of the finite
volume energy shift for two hadron states in a moving frame, confirming the result
obtained by ref.~\cite{rg}.
Our result can be used to determine
the finite-volume corrections to matrix elements of local
composite operators with an initial and/or final state consisting of
two hadrons, thus 
generalizes the Lellouch-L\"uscher factor to moving frames~\cite{kss,cky}.
The path is now open to numerical studies of two pion energies
and matrix elements in a moving frame, which, as discussed
in refs.~\cite{rg,kss,cky}, simplifies a number of
important applications.
\bibliography{proceeding}
\bibliographystyle{JHEP}
\end{document}